# Ulugh Beg, Prince of Stars

Jean-Pierre Luminet

*Aix-Marseille Université, CNRS, Laboratoire d'Astrophysique de Marseille (LAM) UMR 7326
& Centre de Physique Théorique de Marseille (CPT) UMR 7332
& Observatoire de Paris (LUTH) UMR 8102
France
E-mail:* jean-pierre.luminet@lam.fr

## Abstract

Who was Ulugh Beg? A prince who governed a province in the central Asian empire built by his grandfather Tamerlane. Above all, he was a scholar who founded the Samarkand astronomical observatory, whose work predated that of the best astronomers in Europe one and a half centuries later.

## Introduction

In the period between Greek Antiquity and the Renaissance, science in Europe was at a standstill. Arabic astronomy, on the other hand, was marked by doubts about the Greek astronomer Ptolemy's system and by the construction of increasingly accurate instruments, of which the Samarkand observatory is a fine example.

In 1429, Samarkand, a major stopover on the Silk Road, was even livelier than usual. The largest observatory ever built had just been inaugurated. Ambassadors flocked from all over the world to contemplate a 40-metre diameter sextant in a pit 40 metres deep, a gigantic sundial with external walls covered with a fresco depicting the zodiac, and a terrace housing the most advanced instruments for measuring time and space: clepsydra, astrolabes, etc.

The promoter of this architectural prowess was called Ulugh Beg. Director of the observatory, he was also the prince and governor of Samarkand. He was a lover of science and the sky, but a poor politician and soldier - which ultimately caused his death. Ulugh Beg surrounded himself with the best astronomers of the time, observing and calculating the positions of a thousand stars. He wrote a major work: Zij-i Sultani.

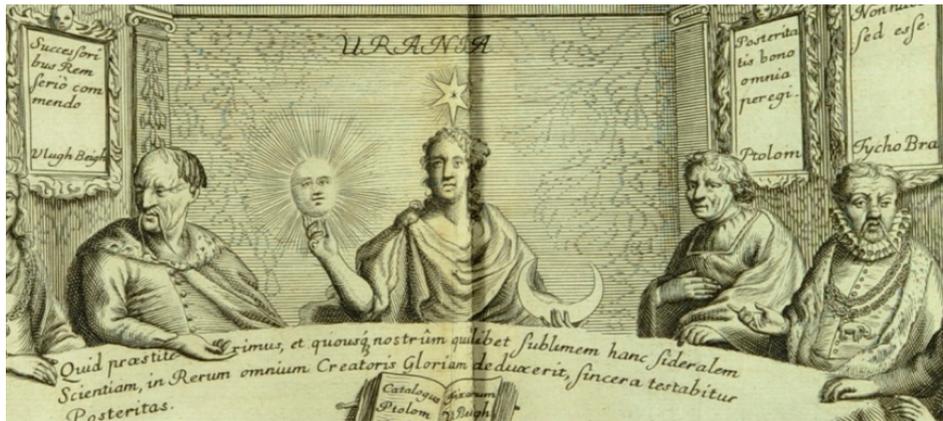
Frontpage of *Prodromus Astronomiae* (1690) by Johannes Hevelius, showing Ulugh Beg sitting near Urania with Ptolemy, Copernicus and Tycho Brahe (detail)

However, this major character in Arabic-Muslim astronomy is all but forgotten in the Western World. Let us do justice to this 15th century Turkish-Mongol prince who reigned over Transoxiana - a huge province of central Asia around its capital Samarkand. Neglecting political affairs in favour of science, he held aloft the torch of Arabic-Muslim astronomy before it was inexorably extinguished by the fires of religious obscurantism.

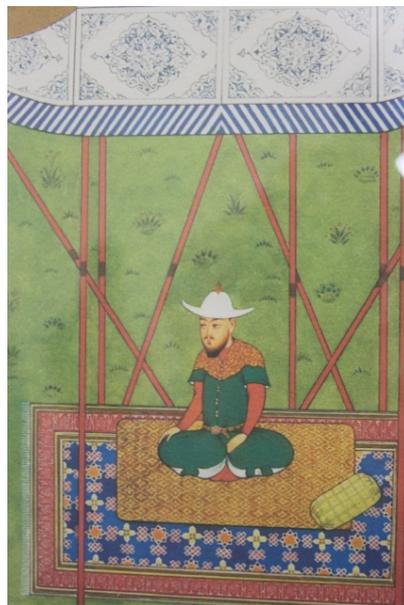
Detail of a Persian miniature depicting Ulugh Beg receiving ambassadors.

## The Conqueror's Grandson

Mohammed Taragae, or Ulugh Beg ("great prince") as he was known, was one of the many grandsons of Timur Leng (1336-1405), better known in the West as Tamerlane. A pitiless conqueror, Tamerlane spread terror during his forty years of reign, building a huge empire in fire and blood which stretched across the current lands of Uzbekistan, Afghanistan, Irak, Iran, Armenia and Georgia. On the margins of the massacres however, Tamerlane gave orders to spare scholars, the educated, artisans and qualified workers so they could be deported to Samarkand to pursue their art. In 1404, just before invading China, Timur died suddenly and his empire was shared among his descendants, leading to fratricidal power struggles. Chah Rukh (1377-1447), the fourth and worthiest son of Tamerlane, ultimately set himself up as sovereign of the largest part of the empire, moving his capital to Herat (in Afghanistan today). He was instrumental in the "Timurid Renaissance", a brilliant but ephermal time when arts, science and culture flourished in the Muslim world.

## A Woeful Warrior

Ulugh Beg was born in Sultaniya (on the Caspian Sea, currently in Iran), on 22 March 1394. He spent most of his childhood living the trials and tribulations of accompanying his grandfather through central Asia as he conquered territories there. When Shah Rukh took power, he appointed his son, hardly 16 years old, governor of Samarkand. Under the benevolent yet severe authority of his father, the young prince Ulugh Beg continued to develop his province of Transoxiana, sending his armies with uncertain success to hold the land together, constantly under threat from a variety of hoards.

Ulugh was a woeful warrior; unlike his grandfather, he was less interested in conquering land than in science and the arts. His education was entrusted to astronomer and mathematician Qadi Zadeh al-Rumi, born in Anatolia in 1364. He had left Turkey, fallen into the hands of Tamerlane, and completed his studies in the Timurid empire's renowned madrasas (a madrasa is a school which served as a university in the Muslim world. The most ancient preserved example, in Ispahan, dates from 1175). Qadi Zadeh knew how to awaken a taste for study and reflection in his pupil, who proved very talented for all the disciplines of the mind, from astronomy to mathematics, including music, poetry and calligraphy. Ulugh Beg had three madrasas built, the largest of which stood in Samarkand's majestic Registan Square. Under his reign, students flocked from all over the East to avail themselves of the teachings of the best professors in theology, literature, sciences or poetry.

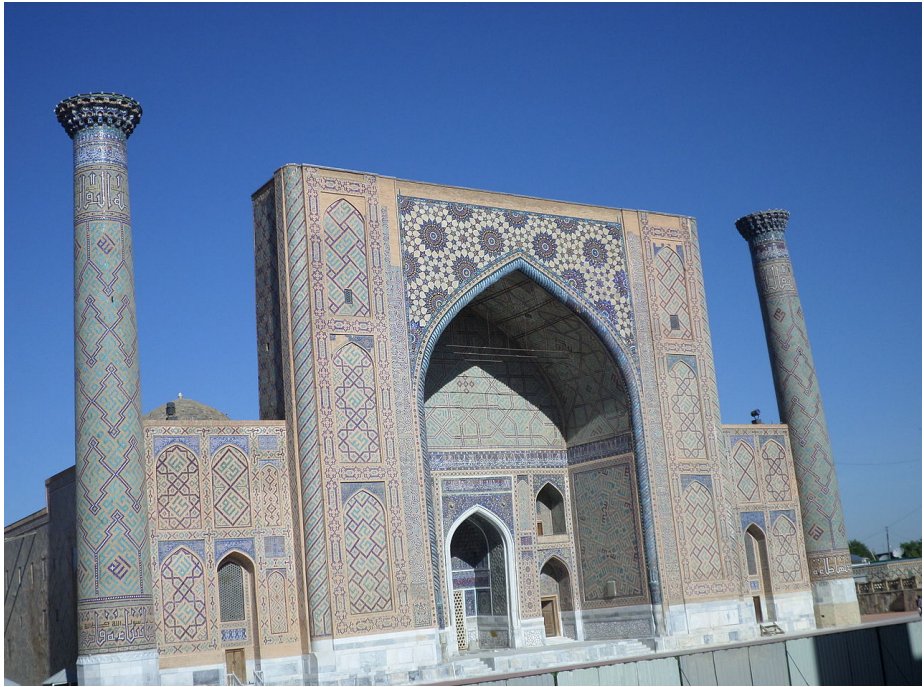

The madrasa built in Samarkand's Registan square under the reign of Ulugh Beg, between 1417 and 1420.

## The Timurid Renaissance

At its height, the Samarkand madrasa housed up to 70 scholars, with the apotheosis being the construction of a giant observatory started in 1424 and inaugurated with great fanfare in 1429.

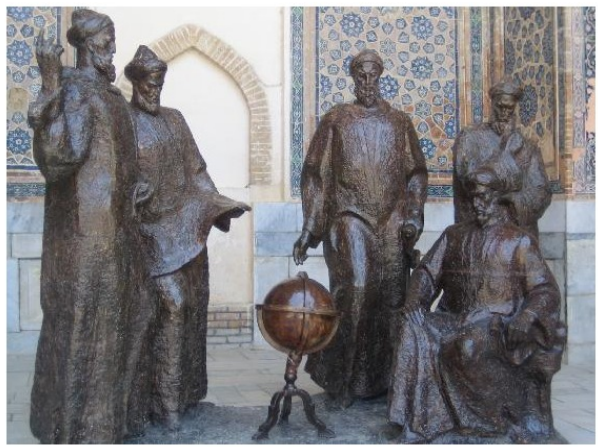

Modern statues representing Ulugh Beg with his favourite astronomers Qadi-Zadeh, Al-Kashi and Ali-Qushji (Samarkand Museum).

Built on the model of the most ancient buildings in Islam such as those in Bagdhad, Ray (near Teheran) and especially Maragha, the Samarkand observatory stood out for its gigantic sextant (an arc comprised of one sixth of a circle). The use of a mural quadrant - a quarter of a circle - to measure the height of stars and planets was very ancient, but unlike previous ones, the Samarkand arc had not one, but two parallel walls 40 meters in radius, double that of the largest instrument of its type ever built previously. The observatory consisted of a cylindrical building three stories high, 48 metres in diameter and reaching a height above ground of 45 metres. It housed many measurement and observation instruments placed on the upper terrace, including a triquetrum, an azimuth circle and an armillary sphere (an instrument that modeled the celestial globe and showed the movement of the stars, sun and the planets around the earth). Its whole structure however, was built around the giant double sextant, aligned with the meridian and used to measure the position of stars above the horizon and their passage across the meridian. The useful part of the instrument was an arc, divided into degrees (from 20 to 80°), minutes and possible angle seconds, limited on each side by two marble-covered edges, on which the 0.698 metre gap between two notches corresponded to one degree of an angle. The lower part of the instrument lay in darkness in an 11-metre deep underground pit.

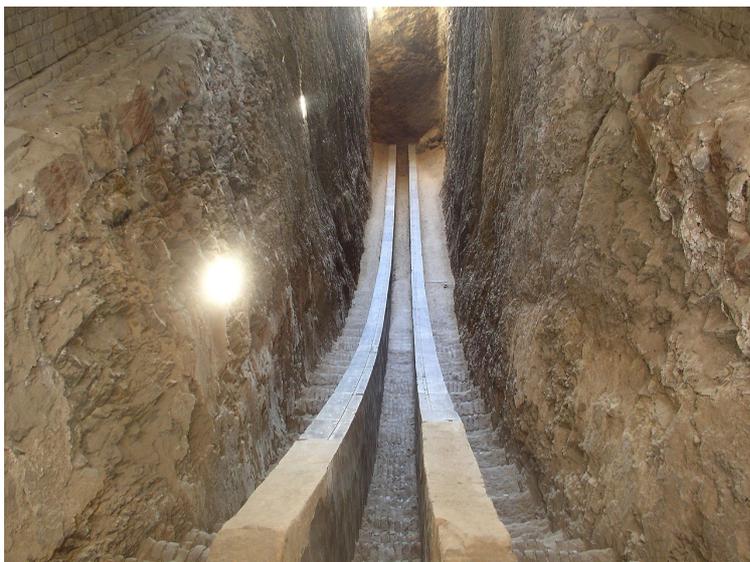
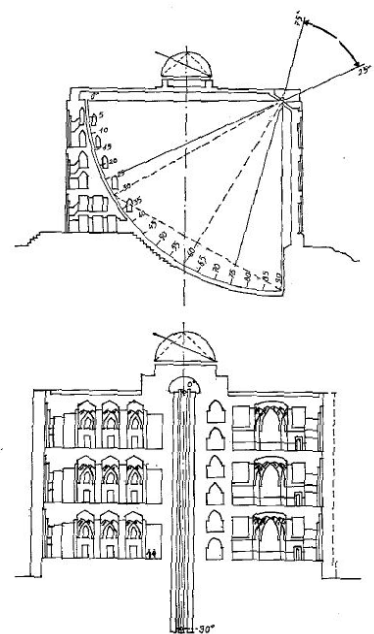

Left : Graduated low part of the double wall sextant. The two arcs are separated by 0,698 meter. The projected image of a star crossing the meridian moved from one wall to the other in exactly four minutes of time, which allowed the instrument to function also as a clock.
Right : Sectional view of the Samarkand Observatory and its built-in quadrant.

One of the most important measurements made by the Samarkand atromers was the obliqueness of the ecliptic, i.e. the angle that the Sun's trajectory plane in the celestial sphere makes with the plane of the equator. It is essential to measure it accurately for astronomical calculations and for the calendar. The value obtained was 23 degrees 30 minutes and 17 seconds, only 32 seconds out from the value recalculated today. The sidereal year, namely the length of time it takes the sun to return to the same position compared to the stars in the celestial globe, is given at 365 days 6 hours 10 minutes and 8 seconds, a difference of 58 seconds (0.04%) with the modern value of 365.25636304 days! Over twenty years, the work of the observatory led to the writing of the Zij-i-Gurgani (Prince's Treatise), a work including data and tables to calculate the position of the sun, the soon and the planets, and a catalogue of 1,018 stars, as accurate as the one established a century and a half later by Tycho Brahe in Europe.

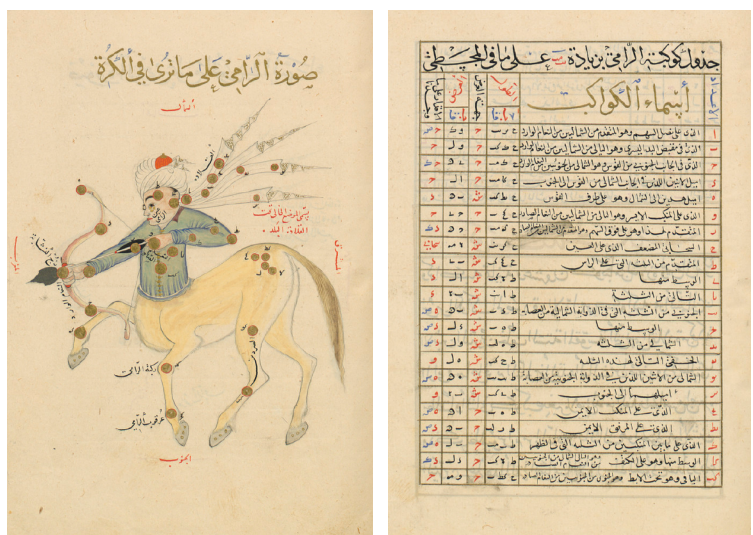

This magnificent reproduction of Al-Sufi's *Book of Fixed Star* belonged to and was annotated by Ulugh Beg. Left, the constellation of Sagittarius. Right, the coordinates and stellar magnitudes revised according to new observations performed at the Smarakand Observatory.

## Influence in Asia

Under the influence of Ulugh Beg, Samarkand extended its influence across the Orient and beyond. But by devoting most of his life to science, the arts and the cultural development of his town, the Transoxiana emir drew the ire of the religious authorities, in particular hat ofhis son Abdulatif, an ambitious fanatic raised far from the court by ulamas (theologians). Ulugh Beg had another son, Abdelaziz, whom he taught astronomy and tolerance, and who he hope to make his successor. In the death of his father Shah Rukh in 1447, Ulugh Beg inherited the destinies of the empire against his will. His reign was short: in 1448, the Uzbeks led by Abdulatif invaded Transoxiana, Samarkand was ransacked, Abdelaziz was killed by his own brother and on 27 October 1449, Ulugh Beg was beheaded on the orders of his eldest son, known in history as "the patricide". The observatory was razed to the ground a few years later and sank into oblivion, until 1908 when a Russian archaeological dig found the ruins, namely the underground pit of the large sextant. The astronomer Ali-Qushji was able to flee Samarkand with his family, taking with him the precious manuscript of astonomical tables. Following an epic journey, he arrived in Constantinople, where he ceremoniously handed the work over to Sultan Mehmet II. He published the works under the title of Sultani Astronomical Tables.

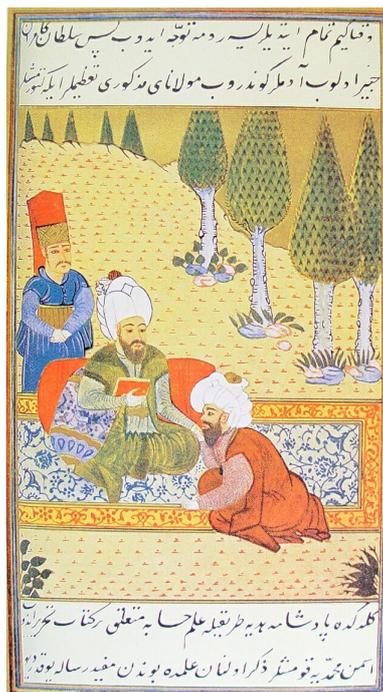

Ottoman miniature showing Sultan Mehmet II receiving the *Sultani Astronomical Tables* from the hands of Ali-Qushji

The influence of the works of Ulugh Beg was felt first in Asia, through he development of Turkish astronomy and the construction of observatories inspired by the one in Samarkand. This was the case in India with the five observatories - all called Jantar Mantar - built at the start of the 18th century by the Maharajah Jai Singh II, a huge admirer of Ulugh Beg. However, the Sultani Astronomical Tables only arrived in Europe in the 17th century. The work of

Ulugh Beg conducted in his observatory was translated and edited for the first time in Oxford in 1648.

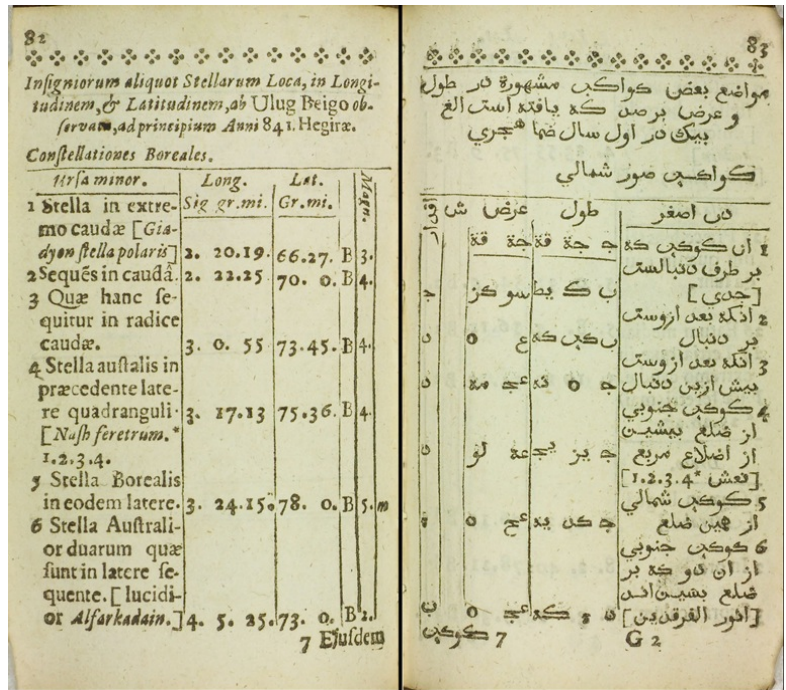
A page of the first western edition of Ulugh Beg's catalog, translated and commented by John Greaves (Oxford, 1648)

## Limited Reception

At the time when Ulugh Beg's work was beginning to be known and disseminated however, Europeans already had more efficient observatories, such as the one set up by Tycho Brahe in Uraniborg in Sweden (a similar type of instrument but with fine graduations enabling more accurate observations), Flamsteed in Greenwich, England, and Hevelius in Dantzig, today Gdansk in Poland. Astronomical instruments had also considerably evolved with the development of optics; the refractor appeared at the start of the 17th century, followed by the telescope. All this and the fabulous theoretical advances of Copernicus, Kepler, Galileo and Newton clearly diminished the influence of Ulugh Beg's work.

Our hero's body was found in 1941 in the Gour Emir, the mausoleum for Timurid sovereigns in Samarkand. Examination of his skeleton revealed a violent blow to the left side of the head, which passed through the lower jaw and cleaved the third cervical vertebra in half. The prince of Samarkand, more passionate about celestial beauty than terrestrial affairs, paid for his love of the stars with his life...